\documentclass[intlimits,twoside,a4paper]{article}
\usepackage{amsmath,amssymb}
\usepackage{graphicx}
\usepackage{wrapfig}
\usepackage[T2A]{fontenc}
\usepackage[cp1251]{inputenc}
\usepackage[eqsecnum]{cmpj}

\issue{2011}{14}{1}{13701}

\doinumber{10.5488/CMP.14.13701}

\title[Dynamic conductivity of symmetric three-barrier plane
nanosystem]%
{Dynamic conductivity of symmetric three-barrier plane nanosystem
in constant electric field}
\author[Ju. Seti, M. Tkach, O. Voitsekhivska]{Ju. Seti, M. Tkach, O. Voitsekhivska}

\address{Fedkovych Chernivtsi National University, 2, Kotsyubinsky St.,
Chernivtsi, 58012, Ukraine\thanks{E-mail: ktf@chnu.edu.ua}}

\date{Received June 16, 2010}
\begin{document}

\maketitle

\begin{abstract}
The theory of dynamic conductivity of nanosystem is developed
within the model of rectangular potentials and different effective
masses of electron in open three-barrier resonance-tunnel
structure in a constant homogeneous electric field.

The application of this theory for the improvement of operating
characteristics of quantum cascade laser active region (for the
experimentally investigated
In$_{0.53}$Ga$_{0.47}$As/In$_{0.52}$Al$_{0.48}$As heterosystem) proves
that for a certain geometric design of nanosystem there exists
such minimal magnitude of constant electric field intensity, at
which the electromagnetic field radiation power together with the
density of current flowing through the separate cascade of quantum
laser becomes maximal.

\keywords resonance-tunnel structure, dynamic conductivity
\pacs 73.21.Fg, 73.90.+f, 72.30.+q, 73.63.Hs
\end{abstract}

\section{Introduction}

Recently, there has been achieved a considerable progress in the
experimental fabrication of quantum cascade lasers
(QCL)~\cite{1,2,3,4} and quantum cascade detectors
(QCD)~\cite{5,6,7,8} of various geometric design. The
investigation of these devices attracts great attention due to
their operation in the actual terahertz range of electromagnetic
waves. The main focus is made on the optimization of parameters of
nano-devices. However, this is a rather complicated problem due to
the absence of a consequent and consistent theory of physical
processes taking place in open nanosystems. The active operating
elements of experimental QCL or QCD are the open resonance-tunnel
structures (RTS) with different number of barriers and wells.
Thus, the properties of their static and dynamic conductivities
determining the basic QCL parameters, i.e., regions and widths of
ranges of operating parts, radiation intensity, excited current
and so on, have been theoretically studied.

In references~\cite{9,10,11,12,13,14,15}, mainly within the model
of unitary effective mass and $\delta$-like potential barriers,
there have been developed the theoretical approaches to the
calculation of active conductivity of electrons in open RTS.
Recently, in references~\cite{16,17} it was shown that
$\delta$-barrier model with unitary electron effective yields too
rough magnitudes of resonance widths of quasistationary states
(ten times bigger) relatively to the realistic model of
rectangular potential barriers with different effective masses of
quasiparticle in different pars of RTS. The conductivity is very
sensitive to the magnitudes of resonance widths of quasistationary
states. Therefore, the rectangular potential barriers and
different effective masses are to be taken into account within the
framework of the respective model. In the majority of theoretical
papers dealing with the conductivity of open RTS, the presence of
constant electric field has not been taken into account at all or
has been evaluated only roughly~\cite{18}. However, the effective
QCL~\cite{1,2,3,4} has been experimentally produced just at the
applied constant electric field. Thus, an urgent task is to
develop a consistent theory of open RTS conductivity at an applied
constant electric field; the model would be deprived of the rough
$\delta$-like approximating barriers and would consider different
effective masses of quasiparticles in the wells and barriers.

In the proposed paper, there is developed a theory of electronic
conductivity of open symmetric three-barrier RTS under the applied
constant homogeneous electric field within the framework of the model
of different electron effective masses in component parts of a
nanosystem with rectangular potential wells and barriers. For the
first time, the obtained exact solutions of the equations
determining the magnitude of the active conductivity at the
applied constant electric field at RTS make it possible to investigate it in
a weak signal one-mode approximation over the radiation field
intensity. By the example of experimental nanosystem
In$_{0.53}$Ga$_{0.47}$As/In$_{0.52}$Al$_{0.48}$As, it is shown that using
the presented model it is possible to optimize the operating
parameters of QCL depending on its geometric design and electric
field intensity.

\section{Hamiltonian. Conductivity of three-barrier nanosystem}

The open symmetric three-barrier RTS with the applied constant
electric field characterized by intensity $F$ (figure~1) is under
study.
\begin{figure}[ht]
\centerline{\includegraphics[width=0.45\textwidth]{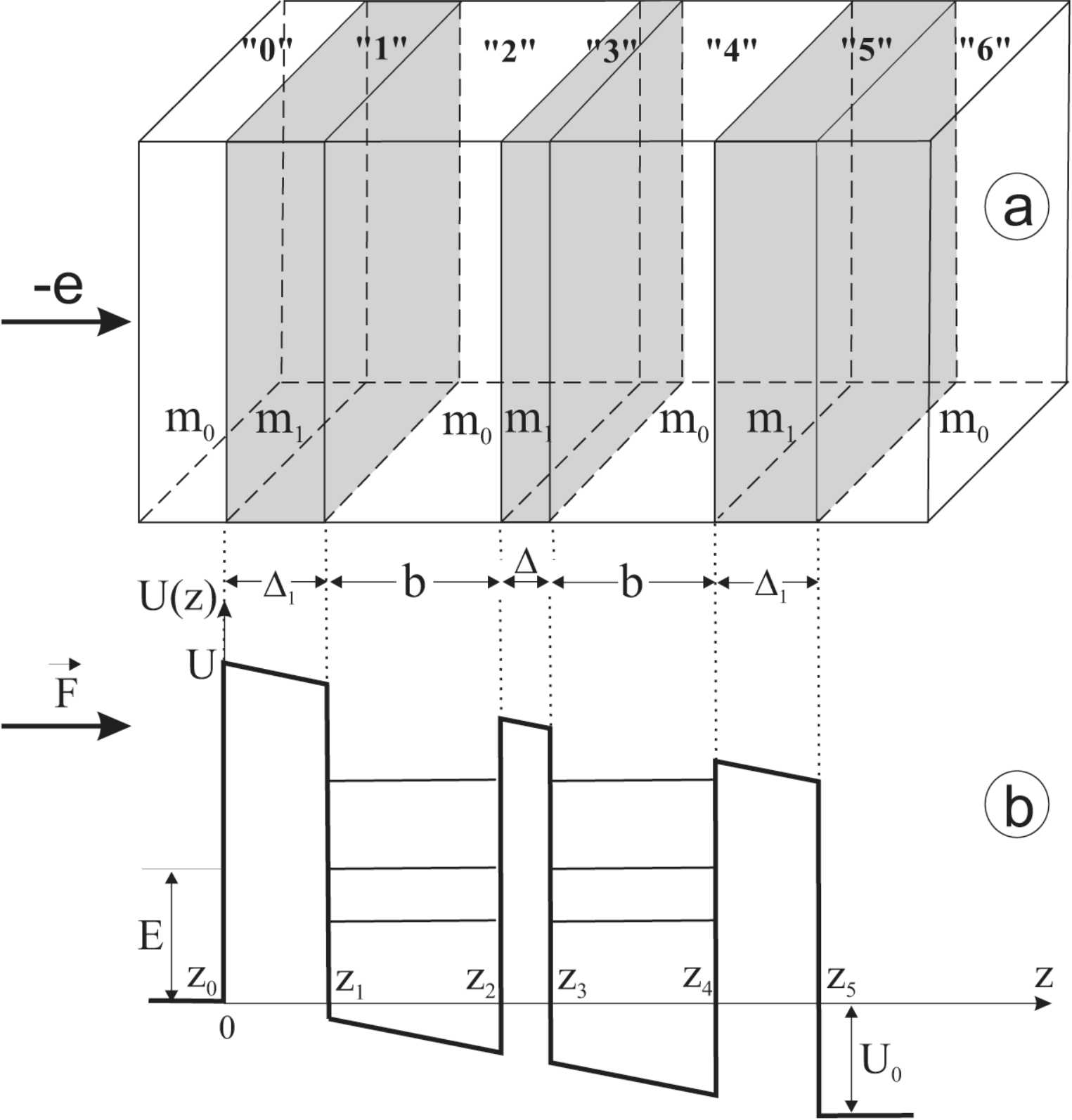}}
\caption{Geometrical (a) and potential energy (b) schemes of RTS.}
\label{fig-smp1}
\end{figure}
It is assumed that the monoenergetic ($E$, the energy)
electron current ($n$, the concentration) is falling at the RTS
from the left side, perpendicularly to its planes. The small
difference between lattice constants of RTS layers-wells (the
media: $0, 2, 4, 6$) and layers-barriers (the media: $1, 3, 5$)
allows us to study the nanosystem within the framework of the model of
effective masses and rectangular potential barriers
\begin{equation}
 \label{eq1} m(z) = {\left\{ {{\begin{array}{*{20}c}
 {m_{0}}  \hfill \\
 {m_{1}}  \hfill \\
\end{array}} } \right.}{\rm ,}
\quad U(z) = {\left\{ {{\begin{array}{*{20}c}
 {0,\,\,\,\,\,\,\,\text{in reg.}\,\,0,\,2,\,4,\,6,} \hfill \\
 {U,\,\,\,\,\,\,\text{in reg.}\,\,1 ,\,3 ,\,\,5. \,\,\,\,\,\,\,\,\,\,} \hfill \\
\end{array}} } \right.}\,\,
\end{equation}

The Schrodinger equation for the electron is written as
\begin{equation}
\label{eq2} \ri\hbar {\frac{{\partial \Psi (z,t)}}{{\partial \,t}}}
= [H_{0} + H(z,t)] \Psi (z,t){\rm ,}
\end{equation}
where
\begin{equation}
\label{eq3} H_{0} = - {\frac{{\hbar ^{2}}}{{2}}}{\frac{{\partial}
}{{\partial z}}}{\frac{{1}}{{m\left( {z}
\right)}}}{\frac{{\partial} }{{\partial z}}} + U(z) - \re F \{z
[\theta(z)-\theta(z-z_{5})]+z_{5} \theta(z-z_{5})\}{\rm }
\end{equation}
-- the Hamiltonian of stationary problem (with constant electric
field),
\begin{equation}
\label{eq4} H(z,t) = - e \epsilon \{z
[\theta(z)-\theta(z-z_{5})]+z_{5} \theta(z-z_{5})\} ( \re^{\ri \omega
t} + \re^{-\ri \omega t} )
\end{equation}
-- the interaction Hamiltonian of electron with the  electromagnetic field varying in
time ($\omega$, the frequency) and amplitude
of electric field intensity ($\epsilon$).

The solution of equation (2.2) in one-mode approximation, assuming
the amplitude of high frequency field to be small
~\cite{12,13,14,18}, according to the perturbation theory, is as
follows:
\begin{equation}
\label{eq5} \Psi \left( {z,t} \right) = \Psi _{0} \left( {z}
\right)\re^{-\ri \omega _{0} t} + \Psi _{ + 1} \left( {z} \right)\re^{-\ri \left( {\omega _{0} + \omega}  \right)t} + \Psi _{ - 1} \left(
{z} \right)\re^{-\ri \left( {\omega _{0} - \omega} \right)t}\,,
\qquad \left( {\omega _{0} = {{E}
\mathord{\left/ {\vphantom {{E} {\hbar} }} \right.
\kern-\nulldelimiterspace} {\hbar} }} \right).
\end{equation}

Here $\Psi _{0}(z)$ function is the solution of stationary
Schrodinger equation
\begin{equation} \label{eq6}
H_{0}(z)\, \Psi _{0} (z)=E\, \Psi _{0} (z).
\end{equation}
Considering that the energy of electronic current in X0Y plane is
negligibly small ($k_{||}=0$) it can be written as follows:
\begin{eqnarray} \label{eq7}
 \Psi _{0} (z)&=&\Psi^{(0)} _{0}(z)\theta(-z)+\sum\limits_{p = 1}^{5}
\Psi^{(p)} _{0}(z) [\theta(z-z_{p-1})-\theta(z-z_{p})]+\Psi^{(6)}
_{0}(z)\theta(z-z_{5}) \nonumber\\
&=&\left(\re^{\ri k^{(0)} z} +B^{(0)} \re^{-\ri k^{(0)} z}\right)\theta(-z)+A^{(6)}
\re^{\ri k^{(6)} (z-z_{5})} \theta (z-z_{5} )\nonumber\\
&&{}+ \sum\limits_{p = 1}^{5} \left[A^{(p)} \mathrm{Ai}(\xi^{(p)})+B^{(p)}
\mathrm{Bi}(\xi^{(p)})\right] \left[\theta(z-z_{p-1})-\theta(z-z_{p})\right],
\end{eqnarray}
where $\mathrm{Ai}(\xi), \mathrm{Bi}(\xi)$ are the Airy functions and
\begin{eqnarray} \label{eq8}
k^{(0)} &=&\hbar ^{-1} \sqrt{\displaystyle 2m_{0} E} , \qquad k^{(6)} =\hbar ^{-1} \sqrt{\displaystyle 2m_{0} \left (E+\re Fz_{5} \right)} ,\nonumber\\
\xi^{(1)}&=&\xi^{(3)}=\xi^{(5)}=\displaystyle \rho^{(1)}
\left[(U-E)/\re F-z \right], \qquad
\rho^{(1)}=\rho^{(3)}=\rho^{(5)}=\displaystyle(2m_{1}\re F/\hbar^{2})^{1/3},\nonumber\\
\xi^{(2)}&=&\xi^{(4)}=-\displaystyle \rho^{(2)} \left(E/\re F+z
\right), \qquad
\rho^{(2)}=\rho^{(4)}=\displaystyle(2m_{0}\re F/ \hbar^{2})^{1/3}
.
\end{eqnarray}

The unknown coefficients ($B^{(0)}, A^{(6)}, A^{(p)}, B^{(p)}\,\
(p=1,\dots ,6)$) are fixed by the fitting conditions for the wave
functions and their densities of currents at all media interfaces
\begin{equation} \label{eq9}
\Psi_{0}^{(p)} (z_{p} )=\Psi _{0}^{(p+1)} (z_{p} ), \quad
\quad \displaystyle \frac{1}{m_{0(1)} } \left. \frac{\rd{\rm \,\ ;
}\Psi _{0}^{(p)} }{\rd z} \right|_{z=z_{p} } =\frac{1}{m_{1(0)} }
\left. \frac{\rd{\rm \; }\Psi _{0}^{(p+1)} (z)}{\rd z} \right|_{z=z_{p}
}, \qquad  (p=0,\dots ,5)
\end{equation}
together with the normalizing condition
\begin{equation} \label{eq10}
\int\limits_{-\infty }^{\infty }\Psi _{0}^{*} (k'z)\Psi _{0} (kz)
{\rm \; }\rd z=\delta (k-k') .
\end{equation}

In order to define the both terms of corrections
($\Psi_{\pm1}(z)$) to wave function, taking into account equations
(2.2)--(2.6) and preserving the magnitudes of the first order,
 inhomogeneous equations are obtained:
\begin{equation} \label{eq11}
[H_{0}(z)-\hbar (\omega_{0}-\omega)]\Psi_{\pm1}(z)-\re \epsilon \{
[\theta(z)-\theta(z-z_{5})]+z_{5} \theta(z-z_{5})\} \Psi_{0}(z)=0
.
\end{equation}
Their solutions are the superposition of functions
\begin{equation} \label{eq12}
\Psi_{\pm1}(z)=\Psi_{\pm}(z)+\Phi_{\pm}(z) ,
\end{equation}
where $\Psi_{\pm}(z)$  are the partial solutions of homogeneous and
$\Phi_{\pm}(z)$ are the partial solutions of inhomogeneous
equations (2.11).

The solutions of homogeneous equations are
\begin{eqnarray} \label{eq13}
\Psi _{\pm} (z)&=&\Psi^{(0)} _{\pm}(z)\theta(-z)+\sum\limits_{p = 1}^{5}
\Psi^{(p)} _{\pm}(z)
[\theta(z-z_{p-1})-\theta(z-z_{p})]+\Psi^{(6)}
_{\pm}(z)\theta(z-z_{5}) \nonumber\\
&=&B^{(0)}_{\pm} \re^{-\ri k^{(0)}_{\pm}
z}\theta(-z)+A^{(6)}_{\pm}
\re^{\ri k^{(6)}_{\pm} (z-z_{5})} \theta (z-z_{5} )\nonumber\\
&&{}+ \sum\limits_{p = 1}^{5} \left[A^{(p)}_{\pm}
\mathrm{Ai}(\xi^{(p)}_{\pm})+B^{(p)}_{\pm} \mathrm{Bi}(\xi^{(p)}_{\pm})\right]
\left[\theta(z-z_{p-1})-\theta(z-z_{p})\right],
\end{eqnarray}
where
\begin{eqnarray} \label{eq14}
k^{(0)}_{\pm} &=&\hbar ^{-1} \sqrt{\displaystyle 2m_{0} (E \pm \hbar\omega)}, \qquad k^{(6)}_{\pm} =\hbar ^{-1} \sqrt{\displaystyle 2m_{0} \left (E\pm \hbar\omega+\re Fz_{5} \right)}, \nonumber\\
\xi^{(1)}_{\pm}&=&\xi^{(3)}_{\pm}=\xi^{(5)}_{\pm}=\displaystyle
\rho^{(1)} \left\{[U-(E\pm \hbar\omega)]/\re F-z \right\},
\qquad \xi^{(2)}_{\pm}=\xi^{(4)}_{\pm}=-\displaystyle \rho^{(2)}
\left[(E\pm \hbar\omega)/\re F+z \right].
\nonumber\\
\end{eqnarray}

The partial solutions of inhomogeneous equations (2.11), as it
is shown, are of an exact analytical form
\begin{eqnarray} \label{eq15}
\Phi_{\pm}(z)\!\!\!&\!=\!&\!\displaystyle \pi \frac{\epsilon}{F} \sum\limits_{p = 1}^{5} \displaystyle \left[\mathrm{Bi}(\xi^{(p)}_{\pm}) \int\limits_{1}^{\xi ^{(p)}} \left(\eta-\rho^{(p)} \frac{U(z)-E}{\re F}\right) \mathrm{Ai}\left(\eta \mp  \rho^{(p)} \frac{\hbar \omega}{\re F}\right)
\Psi_{0}^{(p)}(\eta)\rd\eta \right.\nonumber\\
\!\!\!
&&\left.{}-\displaystyle \mathrm{Ai}(\xi^{(p)}_{\pm}) \!\!\int\limits_{1}^{\xi ^{(p)}}
\!\!\left(\eta-\rho^{(p)} \frac{U(z)-E}{\re F}\right) \!\mathrm{Bi}\left(\eta \mp  \rho^{(p)}
\frac{\hbar \omega}{\re F}\right) \Psi_{0}^{(p)}(\eta)\rd\eta\right]\!\!
[\theta(z-z_{p-1})-\theta(z-z_{p})] \nonumber\\
\!\!\!
&&{}\mp \displaystyle \frac{\re \epsilon z_{5}}{\hbar \omega}
\Psi_{0}^{(6)}(z_{5})\theta(z-z_{5}).
\end{eqnarray}

The conditions for $\Psi(z,t)$  wave function continuity (2.5), at
all nanosystem interfaces, lead to the boundary conditions for the
$\Psi_{\pm1}(z)$
\begin{equation} \label{eq16}
\Psi_{\pm1}^{(p)} (z_{p} )=\Psi _{\pm1}^{(p+1)} (z_{p} ),
\qquad \displaystyle \frac{1}{m_{0(1)} } \left. \frac{\rd{\rm
\; }\Psi _{\pm1}^{(p)} }{\rd z} \right|_{z=z_{p} } =\frac{1}{m_{1(0)}
} \left. \frac{\rd{\rm \,\ ; }\Psi _{\pm1}^{(p+1)} (z)}{\rd z}
\right|_{z=z_{p} }, \qquad (p=0,\dots ,5).
\end{equation}
The solution of the system of inhomogeneous equations (2.16)
defines all unknown coefficients ($B^{(0)}_{\pm}$, $A^{(6)}_{\pm}$,
$A^{(p)}_{\pm}$, $B^{(p)}_{\pm}$). Now, $\Psi_{\pm} (z)$ functions
(2.13) and $\Psi_{\pm1} (z)$ corrections (2.15) are definitively
written, since $\Psi(z,t)$ being the whole wave function, is fixed
too.

Introducing the energy of interaction between the electron and
electromagnetic field, which can be calculated as a sum of
electronic wave energies flowing from both sides of RTS due to the
current, the real part of active conductivity in quasiclassic
approximation~\cite{19} is determined by densities of currents
with the energies
\begin{equation} \label{eq17}
\sigma(\omega)=\displaystyle \frac{\hbar \omega}{2 z_{5} e
\epsilon^{2}} \left\{[j(E+\hbar \omega,z_{5})-j(E-\hbar
\omega,z_{5})]-[j(E+\hbar \omega,0)-j(E-\hbar \omega,0)]\right\}  .
\end{equation}
According to the quantum mechanics, the density of current of
uncoupling electrons with concentration $n$ is related to the
whole wave function
\begin{equation} \label{eq18}
j(z,t)=\displaystyle \frac{\re \hbar n}{2 m(z)} \left[\Psi(z,t)
\frac{\partial}{\partial
z}\Psi^{\ast}(z,t)-\Psi(z,t)\frac{\partial}{\partial z}\Psi(z,t)\right]
.
\end{equation}

Thus, taking into account the expressions (2.5), (2.17) and
(2.18), we obtain the final expression for the real part of
nanosystem active conductivity
\begin{equation} \label{eq19}
\sigma(\omega)=\sigma_{\mathrm{l}}(\omega)+\sigma_{\mathrm{r}}(\omega) ,
\end{equation}
where
\begin{equation} \label{eq20}
\sigma_{\mathrm{l}}(\omega)=\displaystyle \frac{\hbar^{2} \omega n}{2
z_{5} m_{0} \epsilon^{2}} (k^{(0)}_{+}
|B^{(0)}_{+}|^{2}-k^{(0)}_{-} |B^{(0)}_{-}|^{2}) ,
\end{equation}
\begin{equation} \label{eq21}
\sigma_{\mathrm{r}}(\omega)=\displaystyle \frac{\hbar^{2} \omega n}{2 z_{5}
m_{0} \epsilon^{2}} (k^{(6)}_{+} |A^{(6)}_{+}|^{2}-k^{(6)}_{-}
|A^{(6)}_{-}|^{2}) ,
\end{equation}
$\sigma_{\mathrm{r}}(\omega), \sigma_{\mathrm{l}}(\omega)$ are the conductivities,
stipulated by electronic current interacting with electromagnetic
field and flowing out ($\mathrm{r}$ -- right side) and back ($\mathrm{l}$ -- left
side) of nanosystem.

\section{Analysis of the results}

The numeric calculations and analysis of the symmetric
three-barrier RTS conductivity ($\sigma$) and function of the
electric field intensity, $F$ (equal to the magnitude of energy
shift $U_{0}=\re F z_{5}$) were performed for
In$_{0.53}$Ga$_{0.47}$As/In$_{0.52}$Al$_{0.48}$As open nanoheterosystem
with physical parameters:  $m_{0}=0.089 m_{e}$, $m_{1}=0.046
m_{e}$,  $a_{0}=0.5867$~nm, $a_{1}=0.5868$~nm, $U=516$~meV. The
geometrical sizes of wells and barriers were taken in the ranges
of values typical of the experimentally investigated nanosystems
\cite{1,2,3,4,5,6,7,8}. It was assumed that the current of monoenergetic electrons
with concentration $n=10^{16}$~cm$^{-3}$ and energy $E$,
corresponding to the resonance energy of the second
quasistationary state ($E_{n=2}$), falls at the RTS from the left
side.

\begin{figure}[ht]
\centerline{\includegraphics[width=0.9\textwidth]{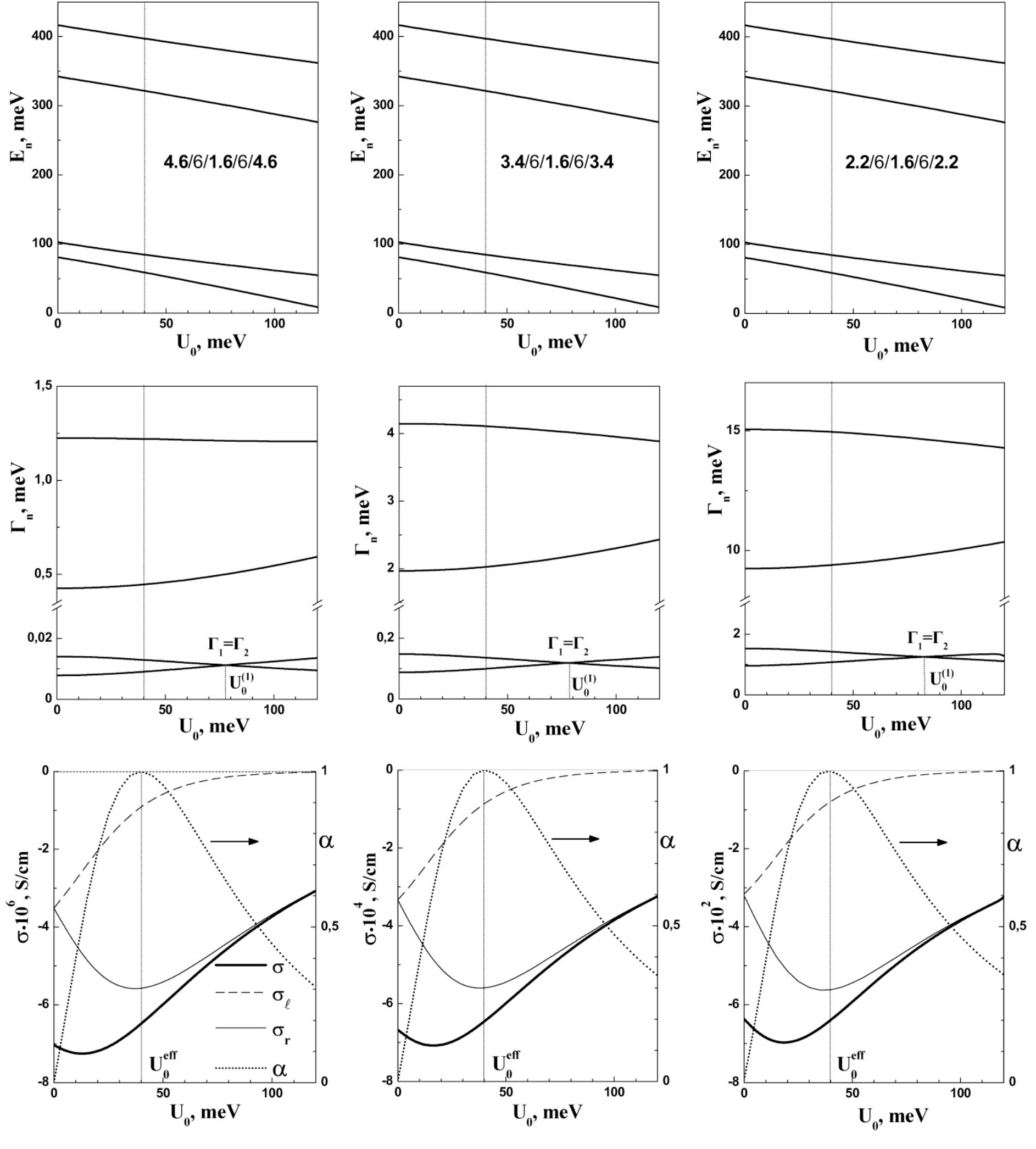}}
\caption{Dependences of resonance energies ($E_{n}$), widths
($\Gamma_{n}$), conductivity ($\sigma$), its terms
($\sigma_{\mathrm{l}}, \sigma_{\mathrm{r}}$) and parameter of
efficiency ($\alpha$) on energy shift ($U_{0}=\re F z_{5}$) and
three different magnitudes the thicknesses of outer barriers at
the fixed sizes of the inner barrier ($\Delta=1.6$~nm) and the
widths of wells ($b=6$~nm).} \label{fig-smp2}
\end{figure}
Figure~2 shows the dependences of resonance energies spectrum
($E_{n}$) and resonance widths ($\Gamma_{n}$) of electron
quasistationary states and conductivity ($\sigma$) at the
transition from the second to the ground quasistationary state and
its terms ($\sigma_{\mathrm{r}}$ -- due to the right-side current and
$\sigma_{\mathrm{l}}$ -- due to the left-side current) on the magnitude
of electromagnetic field shift ($U_{0}$) at the fixed: $b=6$~nm
(wells), $\Delta=1.6$~nm (inner barrier) and $\Delta_{1}=2.2$~nm;
3.4~nm; 4.6~nm (three different magnitudes of outer barriers). In
figure~2: In$_{0.52}$Al$_{0.48}$As barrier layers are in bold,
In$_{0.53}$Ga$_{0.47}$As well layers are in roman. There is also
calculated and presented in the same figure an important
parameter $\alpha=\sigma (\sigma_{\mathrm{r}}-\sigma_{\mathrm{l}})/\mathrm{max}[\sigma
(\sigma_{\mathrm{r}}-\sigma_{\mathrm{l}})]$, characterizing the optimal
efficiency  RTS acting on condition that for the fixed
geometric design of its parts and energy shift ($U_{0}$), both the
conductivity ($\sigma$) proportional to the force of laser
radiation and the magnitude ($\sigma_{\mathrm{r}}-\sigma_{\mathrm{l}}$)
proportional to the  current flowing through the RTS would be
optimal at the same time.

From figure~2 one can see the following. The magnitudes of
resonance energies ($E_{n}$) are almost independent of the
thicknesses of outer barriers ($\Delta_{1}$) at the fixed sizes of
wells ($b$). At the $U_{0}$ increasing (equal to the increasing
intensity of the electric field) all resonance levels show the
linear shift into the region of smaller energies. At the same
time, the resonance widths ($\Gamma_{n}$) of odd states are
increasing and even decreasing. Consequently, there exist such
magnitudes of shifts ($U_{0}^{(p)}$), at which the resonance
widths of neighbouring states are equal. At the increasing
thicknesses of RTS outer barriers, $U_{0}^{(p)}$ magnitudes
do not almost vary and resonance widths exponentially
increase.

According to the developed theory, the intensity of laser
radiation at the frequency $\omega=(E_{2}-E_{1})/\hbar$ would be
almost proportional to the maximal magnitude of negative
conductivity $\sigma$, arising at the transitions of electrons from
the second into the ground quasistationary state. From figure~2 it
is clear that $\sigma$ absolute value firstly increases till
certain maximum value and further linearly decreases for the
bigger $U_{0}$, independently of RTS geometrical parameters. This takes place
 due to the same $\sigma_{\mathrm{r}}$  bahaviuor while
$\sigma_{\mathrm{l}}$ (absolute value)  only decreases at $U_{0}$
increasing.

Figure~2 shows that the parameter of optimal effectiveness
($\alpha$) has one maximum at such $U_{0}^{\mathrm{eff}}$ (for the system
under research, $U_{0}^{\mathrm{eff}}\approx40$~meV relating to the electric
field intensity $F=18$~kVcm$^{-1}; 20 $~kVcm$^{-1}; 22 $~kVcm$^{-1}$ for
the three different sizes of nanosystem) the magnitude of which
almost does not depend on the sizes of outer barriers at the fixed
sizes of wells and outer barriers. It is obvious that
$U_{0}^{\mathrm{eff}}$ magnitude almost coincides with the shift at which
$\sigma_{\mathrm{r}}$ approaches its extremum. Thus, a certain intensity
of constant electric field realizing the shift $U_{0}^{\mathrm{eff}}$ at
which the nanosystem, as an active element of QCL, works optimally,
exists for the fixed geometrical configuration of symmetrical
three-barrier RTS.

As far as the increase of the sizes of the outer barriers causes
an exponential increase of conductivity, as it is clear from
figure~2, when the thicknesses of the outer barriers become bigger
than two lattice constats, the conductivity becomes about  two
orders bigger. Naturally, it does not mean that by increasing the
sizes of outer barriers one can obtain the arbitrarily big
magnitudes of conductivities because it is obvious that herein the
electron lifetime also exponentially increases. For  certain sizes
of nanosystem, the lifetime becomes much bigger than the
relaxation time of electron energy arising due to the
electron-phonon, electron-electron and other types of
interactions. Therefore, the model which does not take into
account these types of energy relaxation becomes useless. Taking
this into consideration we studied such sizes of outer barriers at
which the electron lifetime in the lowest quasistationary states
not bigger than one order exceed the relaxation time due to the
interaction with phonons, evaluated in reference~\cite{1}
approximately as one picosecond.

\section{Conclusions}
\begin{enumerate}
\item  For the first time, there is developed a theory of dynamic
conductivity of symmetric three-barrier RTS with the exact
accounting of the applied constant electric field.
\item It is established that for the fixed geometrical configuration
of RTS there exists one minimal magnitude of electric field
intensity (equivalent to the energy shift $U_{0}$), at which the
active element of QCL works in an optimal regime: the intensity of
electromagnetic radiation and density of current flowing through
the separate cascade of quantum laser become maximal at the same
time.
\item  It is shown that within the framework  of the used models of effective
masses and rectangular potential barriers without taking into
account the electron-phonon, electron-electron interactions and
other relaxation processes, at the fixed sizes of inner barrier
and both wells of three-barrier RTS, the effectiveness of QCL
operation exponentially increases both with the increase of outer
barriers thicknesses as well with the increases of the magnitude
of the shift due to the constant electric field.

\end{enumerate}

\ukrainianpart

\title{─шэрь│ўэр яЁют│фэ│ёЄ№ ёшьхЄЁшўэю┐ ЄЁшсрЁ'║Ёэю┐ яыюёъю┐ эрэюёшёЄхьш
є яюёЄ│щэюьє хыхъЄЁшўэюьє яюы│}
\author{▐.╬.~╤хЄ│, ╠.┬.~╥ърў, ╬.╠.~┬ющЎхї│тё№ър}
\address{╫хЁэ│тхЎ№ъшщ эрЎ│юэры№эшщ єэ│тхЁёшЄхЄ │ьхэ│ ▐Ё│  ╘хф№ъютшўр,  тєы.~╩юЎ■сшэё№ъюую, 2, \\ 58012 ╫хЁэ│тЎ│, ╙ъЁр┐эр}

\makeukrtitle

\begin{abstract}
\tolerance=3000%
╙ ьюфхы│ яЁ ьюъєЄэшї яюЄхэЎ│ры│т │ Ё│чэшї хЇхъЄштэшї ьрё хыхъЄЁюэр
є Ё│чэшї хыхьхэЄрї т│фъЁшЄю┐ ЄЁшсрЁ'║Ёэю┐ Ёхчюэрэёэю-Єєэхы№эю┐
ёЄЁєъЄєЁш, ∙ю чэрїюфшЄ№ё  т яюёЄ│щэюьє юфэюЁ│фэюьє хыхъЄЁшўэюьє
яюы│, ЁючЁюсыхэр ЄхюЁ│  фшэрь│ўэю┐ яЁют│фэюёЄ│ эрэюёшёЄхьш.

\sloppy
╟рёЄюёєтрээ  ЁючЁюсыхэю┐ ЄхюЁ│┐ фы  яюъЁр∙хээ  Ёюсюўшї
їрЁръЄхЁшёЄшъ ръЄштэю┐ юсырёЄ│ ътрэЄютюую ърёърфэюую ырчхЁр (эр
юёэют│ хъёяхЁшьхэЄры№эю фюёы│фцєтрэю┐ ёшёЄхьш 
In$_{0.53}$Ga$_{0.47}$As/In$_{0.52}$Al$_{0.48}$As) яюърчрыю, ∙ю яЁш
чрфрэюьє ухюьхЄЁшўэюьє фшчрщэ│ эрэюёшёЄхьш │ёэє║ Єрър ь│э│ьры№эр
тхышўшэр эряЁєцхэюёЄ│ яюёЄ│щэюую хыхъЄЁшўэюую яюы , яЁш  ъ│щ
юфэюўрёэю ьръёшьры│чє║Є№ё   ъ тхышўшэр яюЄєцэюёЄ│
хыхъЄЁюьруэ│Єэюую тшяЁюь│э■трээ , Єръ │ уєёЄшэр ёЄЁєьє, ∙ю
яЁюїюфшЄ№ ъЁ│ч№ юъЁхьшщ ърёърф ътрэЄютюую ърёърфэюую ырчхЁр.
\keywords Ёхчюэрэёэю-Єєэхы№эр ёЄЁхъЄєЁр, фшэрь│ўэр яЁют│фэ│ёЄ№

\end{abstract}

\end{document}